# Ultrabroadband single-cycle terahertz pulses with peak fields of 300 kV cm$^{-1}$ from a metallic spintronic emitter

T. Seifert[1], S. Jaiswal[2,3], M. Sajadi[1], G. Jakob[2], S. Winnerl[4], M. Wolf[1], M. Kläui[2], T. Kampfrath[1,*]

1. Department of Physical Chemistry, Fritz Haber Institute of the Max Planck Society, 14195 Berlin, Germany
2. Institute of Physics, Johannes Gutenberg University Mainz, 55099 Mainz, Germany
3. Singulus Technologies AG, 63796 Kahl am Main, Germany
4. Helmholtz-Zentrum Dresden-Rossendorf, 01328 Dresden, Germany

* E-mail: kampfrath@fhi-berlin.mpg.de

To explore the capabilities of metallic spintronic thin-film stacks as a source of intense and broadband terahertz electromagnetic fields, we excite a W/CoFeB/Pt trilayer on a large-area glass substrate (diameter of 7.5 cm) by a femtosecond laser pulse (energy 5.5 mJ, duration 40 fs, wavelength 800 nm). After focusing, the emitted terahertz pulse is measured to have a duration of 230 fs, a peak field of 300 kV cm$^{-1}$ and an energy of 5 nJ. In particular, the waveform exhibits a gapless spectrum extending from 1 to 10 THz at 10% of amplitude maximum, thereby facilitating nonlinear control over matter in this difficult-to-reach frequency range and on the sub-picosecond time scale.

Terahertz (THz) pulses covering the range from 1 to 20 THz are useful resonant probes of numerous low-energy excitations in all phases of matter. Completely new research avenues open up when THz pulses are used to drive rather than probe materials resonances.[1,2,3,4] In solids, examples are the ultrafast coherent control over the motion of lattice ions and ordered electron spins, and the transport of charge carriers, even across the atomic-scale junction of scanning tunneling microscopes.[5] To implement such material control, elevated field strengths >100 kV cm$^{-1}$ over a wide frequency range are required. Furthermore, to access more resonances with better time resolution, higher bandwidth is highly desirable.

State-of-the-art strong-field table-top sources[6] at the low-frequency side (0…5 THz) are based on optical rectification in photoconductive switches[7], inorganic[8] and organic[9,10,11] crystals. For frequencies above 10 THz, difference frequency mixing of the two outputs of a dual optical parametric amplifier in a GaSe crystal was shown to yield field strength exceeding 100 MV cm$^{-1}$ (Ref. 12). Regardless of the very high conversion efficiencies reached with these approaches, they are affected by spectral gaps in between 1 and 10 THz. Emission from dual-color-laser-excited air plasma can cover frequencies from <1 to >10 THz with very high field strength.[13] However, the experimental realization is not as straightforward as with emitters relying on optical rectification. Thus, the frequency range from about 5 to 15 THz still poses a challenge in terms of high field sources.[14]

Recently, THz sources based on metallic spintronic heterostructures were shown to be a new and promising class of THz emitters.[15,16,17,18,19,20,21] When excited with 10 fs nanojoule-class laser pulses from an oscillator, a gapless spectrum from 1 to 30 THz and a conversion efficiency comparable to or even better than standard oscillator-based THz emitters were achieved.[22] Advantages of this emitter concept are ease-of-use, stability and a collinear geometry. However, the capability of spintronic THz emitters as high-field sources driven by millijoule-class laser pulses remains to be shown.

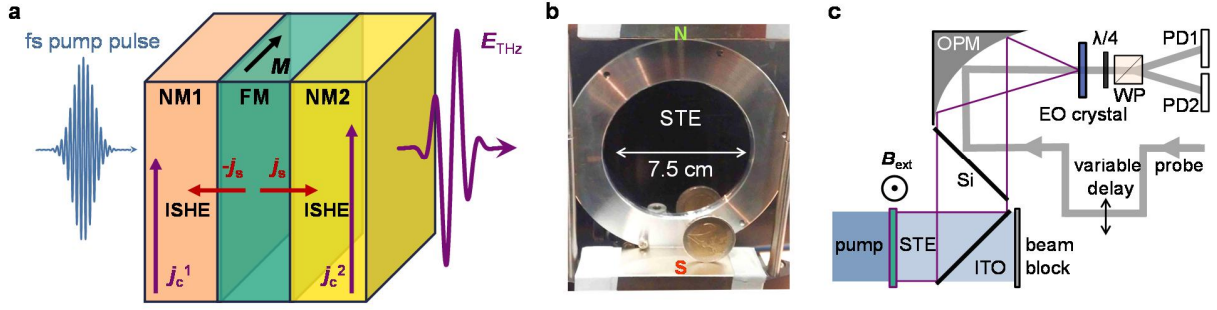

**FIG. 1.** High-field spintronic terahertz emitter (STE). (a) Principle of operation. A femtosecond laser pulse drives spin currents $j_s$ from a ferromagnetic (FM) layer (magnetization $M$ is in-plane) into two adjacent non-magnetic (NM) layers. The inverse spin Hall effect (ISHE) converts these spin currents into orthogonal in-plane charge currents $j_c^1$ and $j_c^2$. By design, NM1 and NM2 have opposite spin Hall angles, thereby resulting in constructive superposition of the two sub-picosecond charge currents. Consequently, a THz pulse $E_{THz}$ is emitted into the optical far-field. (b) Photograph of the spintronic terahertz emitter. Two bar magnets provide a magnetic field of ≥10 mT across the entire emitter area. A 2 €coin serves as a scale reference. (c) Schematic of the experimental setup. For details, see the main text. Abbreviations: $B_{ext}$: external magnetic field, ITO: Indium-Tin-Oxide-covered glass, Si: Silicon wafer, OPM: off-axis parabolic mirror, EO: electrooptic, λ/4: quarter-wave plate, WP: Wollaston prism, PD: photodiode.

Here, we demonstrate upscaling of metallic spintronic THz emitters, resulting in a practical, cheap and ultrabroadband source delivering THz pulses with a duration of 230 fs, a spectrum from 1 to 10 THz (full width at 10% of amplitude maximum) and peak fields of 300 kV cm$^{-1}$.

Our metallic spintronic THz emitter[22] (STE, see Figs. 1a and 1b) is a nanometer-thick trilayer structure NM1/FM/NM2, made of a ferromagnetic (FM) layer FM=$Co_{20}Fe_{60}B_{20}$ between two non-magnetic (NM) layers NM1=Pt and NM2=W on a fused-silica substrate. The detailed stack structure is fused silica (thickness of 500 µm) | W(1.8 nm) | $Co_{20}Fe_{60}B_{20}$(2 nm) | Pt(1.8 nm) (see Supplementary Material). It is worth noting that the cost of the emitter is mainly determined by the substrate price of ~$20.

Upon excitation with a near-infrared femtosecond pump pulse, a distribution of non-equilibrium electrons is created in the emitter. Importantly, the transport properties of the majority and minority spins in the FM layer (i.e. spin-dependent lifetimes, densities and group velocities) differ distinctly. Consequently, and in analogy to the spin-dependent Seebeck effect (SDSE)[23], spin currents polarized parallel to the sample magnetization are injected from the FM into the two adjacent NM layers where spin-orbit coupling causes a spin-dependent deflection of the electrons. This inverse spin Hall effect (ISHE) transforms the spin current into a sub-picosecond transverse charge current[24] that gives rise to the emission of a THz electromagnetic pulse. NM materials showing a particularly large ISHE, yet with opposite sign of the spin Hall angle, are Pt and W.

In our experiment (see Fig. 1c for a schematic), we use laser pulses (energy of 5.5 mJ, center wavelength of 800 nm, duration of 40 fs, repetition rate of 1 kHz) from an amplified Ti:sapphire laser system (Coherent Legend Elite Duo). The collimated beam (diameter of 4.8 cm full width at half intensity maximum, FWHM) is incident onto the STE, whose magnetization is saturated in the sample plane by an array of permanent magnets (field of ±10 mT). To spectrally separate the pump from the THz radiation, the emitted THz beam is reflected by an ITO-coated float glass (thickness of 100 nm, sheet resistance <7 Ω/sq, covered with a $SiO_2$ passivation layer) under an angle of 45°. After transmission through a Silicon wafer (angle of incidence of 45°±2°), which blocks the residual pump radiation, the THz beam is eventually focused on two different detectors that serve to characterize the THz power and the transient THz electric field. The THz power is measured with a power meter (Gentec THz-B), which requires chopping of the 800 nm pump beam at 25 Hz. To determine the polarization state of the THz radiation, we employ a rotatable free-standing wire grid polarizer (InfraSpecs model P02) placed directly behind the Silicon wafer.

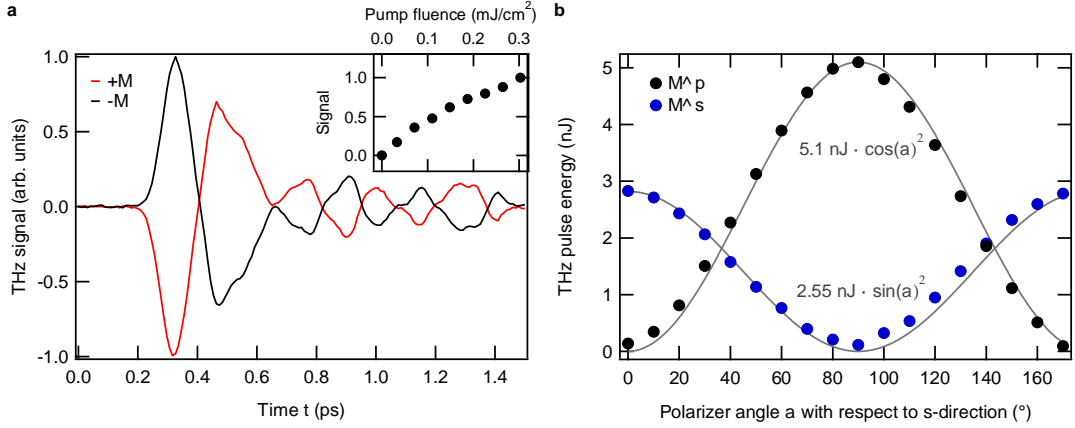

**FIG. 2.** Raw data. (a) Typical THz electrooptic signals measured with a 50 µm thick Quartz detector for opposite sample magnetizations $\pm M$. The inset shows the pump-fluence dependence of the THz signal (RMS). (b) THz pulse energy as a function of the rotation angle $\alpha$ of a THz polarizer inserted before the THz power meter for two orthogonal sample magnetizations (black and blue dots). Grey lines are $\cos^2\alpha$ and $\sin^2\alpha$ fits. THz pulse energies are corrected for polarizer transmission losses.

We characterize the transient electric field of the THz pulse by standard EO sampling using a femtosecond probe pulse from the seed oscillator (energy of 0.6 nJ, center wavelength of 750 nm, duration of 8 fs, repetition rate of 80 MHz) that is coupled into the THz beam path upon reflection from the rear side of the Si wafer.[25] The THz and the near-infrared probe beam are focused by a 45° off-axis parabolic mirror (focal length of 2 inch) into the detection crystal of either (110)-oriented GaP (thickness of 50 µm), (110)-oriented ZnTe (10 and 50 µm) or (001)-oriented Quartz (50 µm). The THz-field-induced probe ellipticity is measured by an optical bridge consisting of a quarter-wave plate, a Wollaston prism and two fast photodiodes. The detection crystals are sufficiently thin to ensure a linear scaling of the EO signal with THz electric field. If not mentioned otherwise, measurements are conducted at room temperature in air. Details on EO detection with a Quartz crystal will be published elsewhere.[26]

Figure 2 displays typical THz signals for opposite magnetization directions as recorded with the Quartz detector. It also shows the dependence on the pump fluence, measurements of the THz polarization state and of the THz pulse energy. Typical THz signals measured with different EO detectors are shown in Fig. 3a. In the following, the data are described and discussed in detail.

Figure 2a shows a typical EO signal as recorded with a 50 µm thick Quartz crystal. We observe an almost complete reversal of the THz signal when the sample magnetization is reversed. This behavior is consistent with our understanding of the THz emission process (see above, Fig. 1a and Ref. 22). The pump fluence dependence demonstrates that the THz emission is still well below saturation (inset of Fig. 2a). We note that the temporal shape of the THz pulse is independent of the pump fluence (not shown). The observed ringing after the main pulse (Fig. 2a) may arise from the THz absorption of water vapour in air. This notion is corroborated by the fact that additional purging with $N_2$ leads to a 20% increase in THz intensity and a slight reduction of the ringing. Another possible origin is the absorption by phonon resonances of the EO Quartz detector at around 4, 8 and 12 THz.[27,28]

Figure 2b displays the measured THz power behind a wire-grid polarizer as a function of its azimuthal rotation angle $\alpha$ for the sample magnetization set parallel (p) and perpendicular

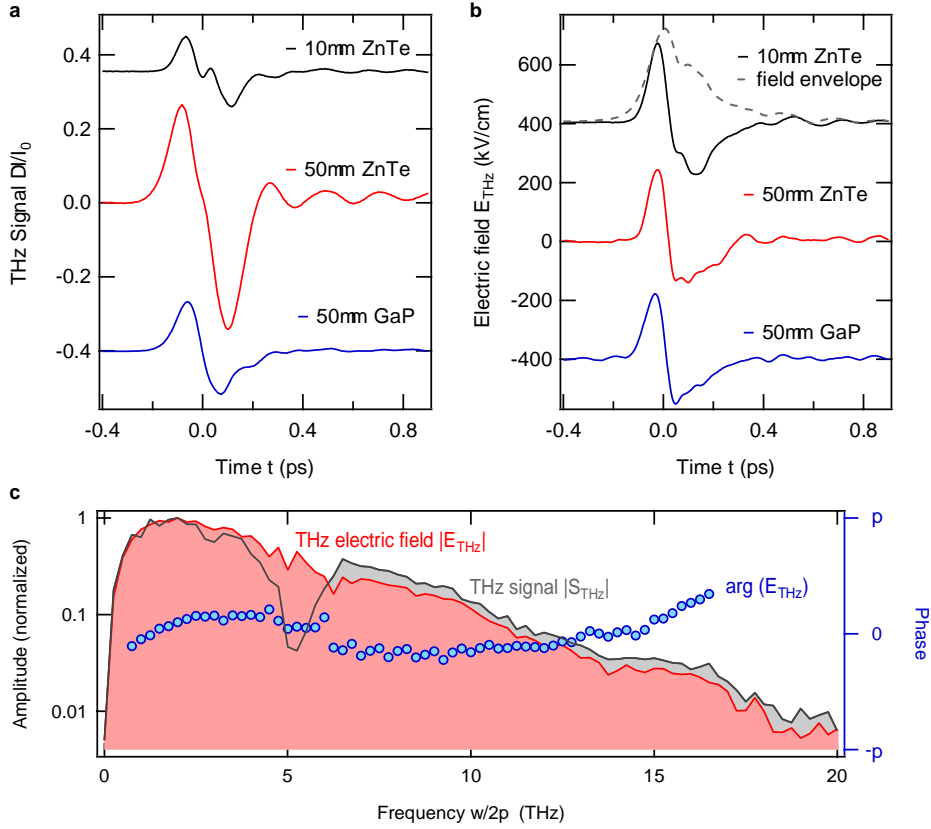

**FIG. 3.** THz-electric-field extraction. (a) Electrooptic signals as recorded with three different detection crystals (10 μm ZnTe, 50 μm ZnTe and 50 μm GaP) in a dry nitrogen atmosphere. (b) Resulting THz electric fields at the detector position obtained by deconvolution of the detector response function. The grey dashed line is the field envelope function. (c) Spectra of signal amplitude, electric-field amplitude and field phase as obtained with the 10 μm thick ZnTe detector.

(s) to the optical table. At $\alpha = 90°$, the signal has a maximum for the $\hat{M} \parallel \mathbf{p}$-configuration whereas a minimum is found for the $\hat{M} \parallel \mathbf{s}$-configuration. The inverse behavior is found at $\alpha = 0°$. The measured data are well described by an $\alpha$-dependence following $\cos^2 \alpha$ and $\sin^2 \alpha$. Therefore, the THz radiation measured by the power detector is polarized linearly and oriented perpendicularly to the sample magnetization. These polarization properties agree with the SDSE/ISHE THz emission scenario of Fig. 1a. The different maximum power amplitudes obtained for the $\hat{M} \parallel \mathbf{p}$- and $\hat{M} \parallel \mathbf{s}$-configurations can easily be explained by the polarization-dependent transmittance of the Si window located between STE and polarizer (see Fig. 1c): a straightforward calculation of the Fresnel transmission coefficients shows that the transmittance ratio of p- and s-polarized THz radiation is $2.0 \pm 0.1$ (Ref. 29). This value is in good agreement with the experimental observation (see Fig. 2b).

Taking the transmission of the polarizer into account (the THz-spectrum-weighted power transmittance is 86%),[30] we obtain a THz pulse energy of about 5.1 nJ (for p-polarization). We note that both coherent THz pulses and incoherent black-body radiation of the pump-heated sample can contribute to the power that is emitted by the STE. Since the black-body radiation is largely unaffected by the magnetization direction of the sample, the power behind the polarizer should be identical for the $\hat{M} \parallel \mathbf{p}$- and $\hat{M} \parallel \mathbf{s}$-configurations. This behavior is, however, not observed (see Fig. 2b). Therefore, the black-body radiation arriving at the power detector gives a minor or probably even negligible contribution to the detector signal. This observation can be explained by the following two scenarios.

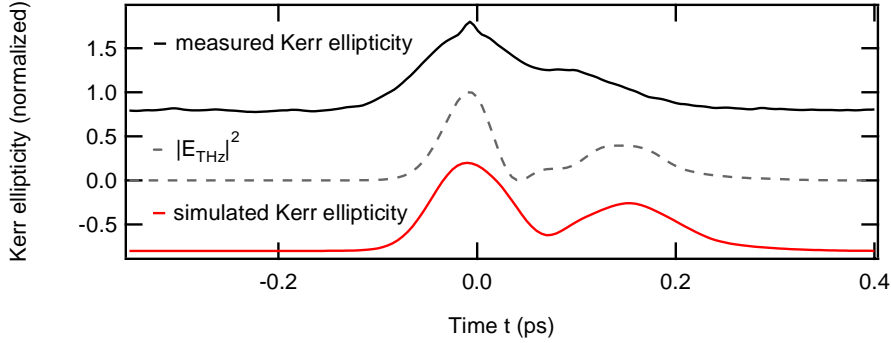

**FIG. 4. Terahertz Kerr effect in diamond.** Transient ellipticity (solid black) of a near-infrared probe pulse induced by a THz pump pulse from the STE. The squared THz electric field (dashed grey) and the simulated Kerr ellipticity (solid black) are also shown. For details of the model see the main text. All curves are normalized to 1.

First, the black-body radiation arriving at the detector has a much smaller power than the coherent THz radiation. Second, the instantaneous STE temperature and the resulting black-body radiation are not able to follow the pump-power modulation frequency of 25 Hz due to the mechanical chopper, thereby being suppressed by our phase-locked power detector. Therefore, each THz pulse has an energy of 5.1 nJ and linear polarization perpendicular to the sample magnetization

To extract the actual THz electric field at the detector position, the measured EO signal is deconvoluted with respect to the transfer function of the EO detection crystal.[25] The deconvolution is performed in the time domain for three different detector crystals (10 µm ZnTe on a (100)-oriented ZnTe substrate[25], 50 µm free-standing ZnTe and GaP). The resulting field waveforms are low-pass filtered with a Gaussian function centered at 0 THz and having a very wide bandwidth of 40 THz FWHM.

Figure 3a shows typical EO signals recorded in a dry Nitrogen atmosphere with the three different detectors. The THz signals are scaled by the intensity ratio $(I_1 - I_2)/(I_1 + I_2)$ measured directly on the two photo diodes of the optical bridge. The respective THz electric fields $E_{THz}(t)$ are presented in Fig. 3b.

We find single-cycle waveforms whose temporal shape and amplitude are in excellent agreement for all the three detectors used. This fact demonstrates the robustness of our deconvolution scheme. The extracted transient THz electric field reaches a peak value of 300 kV cm$^{-1}$ and has a duration of 230 fs (FWHM of the field envelope, see Fig. 3b).

Fourier-transformation of the field waveforms $E_{THz}(t)$ yields the complex-valued field amplitude spectrum that is shown in Fig. 3c along with the respective THz signal spectrum. Note that the THz field spectrum is gapless and spans the entire range from 0.1 to 10 THz (with respect to 10% of the peak spectral amplitude). The spectral phase is flat and varies by less than $2\pi/10$ (standard deviation).

As an important cross-check, we compare the extracted transient THz electric field in the focus (Fig. 3b) to the measured THz energy (Fig. 2b) which are related by:

$$W = C \int_{-\infty}^{\infty} d\omega \ |E(\omega)|^2/\omega^2 \quad (Eq. 1)$$

where $C = 2\pi \ln 2 \, Z_0^{-1} c^2 f^2 b^{-2}$, $Z_0 \approx 377 \, \Omega$ is the free-space impedance, $c$ is the vacuum speed of light, $f = 5.1$ cm is the focal length of the parabolic mirror and $b = 2.4$ cm is the beam radius at half intensity maximum (see Supplementary Material for details). Using the measured THz electric field in the focus (Fig. 3b) and Eq. (1), we obtain a THz pulse energy of 4.1 nJ, which is in excellent agreement with the directly measured value of 5.1 nJ.

To provide a first demonstration of the capability of these pulses for THz nonlinear optics, we measure the THz Kerr effect[31,32,34,33] of diamond. To study this $\chi^{(3)}$-type nonlinear optical effect, the p-polarized THz transient is focused into a 320 µm thick polycrystalline diamond crystal. We measure the transient birefringence using a co-propagating probe beam (same pulse specifications as in EO sampling) linearly polarized with an angle of 45° with respect to the THz field direction. The experiment is conducted in a dry Nitrogen atmosphere.

Figure 4 shows the induced ellipticity acquired with a moderate measurement time of 5 min. Its striking similarity to the squared THz electric field suggests the sample response to be quadratic in the THz field, that is, of $\chi^{(3)}$-type. To support this understanding, we simulate the Kerr-type pump-probe signal by taking the velocity mismatch between pump and probe beam into account.[34] As seen in Fig. 4, we find good agreement with the measured data. Small discrepancies may originate from neglecting lensing effects due to the sharply focused THz field[35] and the dispersion of diamond's THz refractive index. The THz Kerr effect observed here demonstrates the capability of the STE as a high-field THz source.

In conclusion, a large-area spintronic emitter is successfully implemented as a high-field THz source. Excitation by 5.5 mJ optical pump pulses results in single-cycle THz pulses with a duration of only 230 fs and peak electric fields of 300 kV cm$^{-1}$. The capability of these THz pulses in terms of driving non-linear effects is demonstrated by inducing a transient $\chi^{(3)}$-response in diamond. We note that the THz generation mechanism relies on ultrafast electron heating and should, therefore, be virtually independent of the pump wavelength. The combination of ease-of-use, low-cost (substrate price of ~$20), versatility and scalability makes this high-field emitter concept very interesting for THz nonlinear optics. It holds the promise for an even improved emitter performance in the near future.

We emphasize that this work is only a first step toward spintronic strong-field THz sources. Numerous improvements of the currently used emitter are anticipated such as optimization of the fluence and duration of the pump pulse. Finally, in terms of the emitter itself, a number of degrees of freedom can be tuned, including the emitter temperature, the choice of materials with large spin Hall angle[36], the layer sequence and the arrangement of cascaded emitters.[37]

See supplementary material for details on the sample preparation and on the calculation of the pulse energy.

T.S., M.S., M.W. and T.K. acknowledge the European Research Council for funding through the ERC H2020 CoG project TERAMAG/grant no. 681917 and the German Research Foundation for funding through Grant No. KA 3305/2-1 and the priority program SPP 1538/SpinCaT. S.J., G.J., and M.K. acknowledge funding by the European Community under the Marie-Curie FP7 program, ITN "WALL" (Grant No. 608031), the German Research Foundation (in particular SFB TRR173 Spin+X) and the Graduate School of Excellence Materials Science in Mainz (MAINZ).